\documentclass[twocolumn,showpacs,preprintnumbers,superscriptaddress,amsmath,amssymb,nofig]{revtex4-1}
\usepackage[parfill]{parskip}     
\usepackage{amssymb}
\usepackage{amsmath}
\usepackage{amssymb}
\usepackage{amsfonts}
\usepackage{dsfont}
\usepackage[section]{placeins}
\usepackage{bm}
\usepackage{graphicx}
\usepackage{lipsum}
\usepackage{color}
\newcommand{\ket}[1]{|#1\rangle}             
\newcommand{\qo}[1]{``#1''}                     
\usepackage{soul}
\definecolor{lightblue}{RGB}{185,210,248}
\sethlcolor{lightblue}

\begin{document}
\title{Generation of a spin-polarized electron beam by multipoles magnetic fields}
\author{Ebrahim Karimi}
\email{Corresponding author: ekarimi@uottawa.ca}
\affiliation{Department of Physics, University of Ottawa, 150 Louis Pasteur, Ottawa, Ontario, K1N 6N5 Canada}
\author{Vincenzo Grillo}
\affiliation{CNR-Istituto Nanoscienze, Centro S3, Via G Campi 213/a, I-41125 Modena, Italy}
\author{Robert W. Boyd}
\affiliation{Department of Physics, University of Ottawa, 150 Louis Pasteur, Ottawa, Ontario, K1N 6N5 Canada}
\affiliation{Institute of Optics, University of Rochester, Rochester, New York, 14627, USA}
%
\author{Enrico Santamato}
\affiliation{Dipartimento di Scienze Fisiche, Universit\`{a} di Napoli ``Federico II'', Compl.\ Univ.\ di Monte S. Angelo, 80126 Napoli, Italy}
\affiliation{Consorzio Nazionale Interuniversitario per le Scienze Fisiche della Materia, Napoli}
\begin{abstract}
The propagation of an electron beam in the presence of transverse magnetic fields possessing integer topological charges is presented. The spin--magnetic interaction introduces a nonuniform spin precession of the electrons that gains a space-variant geometrical phase in the transverse plane proportional to the field's topological charge, whose handedness depends on the input electron's spin state. A combination of our proposed device with an electron orbital angular momentum sorter can be utilized as a spin-filter of electron beams in a mid-energy range. We examine these two different configurations of a partial spin-filter generator numerically. The results of these analysis could prove useful in the design of improved electron microscope.
\end{abstract}
\maketitle
\section{Introduction}
A few years ago, the existence of orbital angular momentum (OAM) for an electron beam was predicted theoretically~\cite{bliokh07}. A couple of years later, two different techniques, based on the holography and random phase changes in a graphite sheet, were used to generate electron beams carrying OAM experimentally~\cite{uchida10,verbeeck10,mcmorran11}. Such an intriguing topic is of particular interest to materials scientists as it opens up new opportunities for their community~\cite{verbeeck11}. The OAM as a \qo{\textit{rotational-like}} degree of freedom of an electron beam induces a \textit{magnetic moment} in addition to the spin magnetic moment, even up to few hundred Bohr magneton per electron, which gives a possibility to interact with an external magnetic field~\cite{mcmorran11,verbeeck11}. The interaction of OAM magnetic moments with a uniform longitudinal magnetic fields or a fluxes have been recently examined theoretically and experimentally~\cite{bliokh12,greenshields12,guzzinati12}. Indeed, this interaction would enhance or diminish the beam's kinetic OAM, which eventuates in an additional opportunity to measure or sort electron's OAM spatially. However, beside its interesting and fascinated applications, this novel degree of freedom of electrons would be utilized to rock some fundamental quantum concepts such as the \textit{Bohr-Pauli impossibility of generating a spin-polarized free electron beam}~\cite{darrigol84,karimi12}. \newline
The spin-orbit coupling in a non-uniform balanced electric-magnetic field, named a ``$q$-filter'', was proposed by some of the authors as a novel tool to generate an electron vortex beam from a pure spin-polarized electron beam. In that configuration, the spin of an electron follows the \textit{Larmor precession} up and acquires a geometrical phase, which depends on both spin--magnetic field direction and the time of interaction as well. A non-uniform magnetic field introduces a non-uniform phase structure the same as the topological structure of the magnetic field. Several different topological charge configurations, proposed in the previous article, can be manufactured practically~\cite{karimi12}. A local orthogonal electric field was proposed to compensate the \qo{net} magnetic force. Furthermore, the reverse process was suggested to filter the spin component of an electron beam spatially, where two different longitudinal electron's spin components suffer opposite precession directions, thus possess opposite OAM values. \newline

In this work, we suggest a scheme based on \textit{non-uniform magnetic fields}, rather than a balanced space-variant \textit{Wien} $q$-filter, to manipulate electron OAM. As the $q$-filter the proposed scheme imprints onto the incoming electron beam the magnetic field topological charge, with a handedness depending on the longitudinal component of electron spin, positive for spin up $\ket{\!\uparrow}$ and negative for spin down $\ket{\!\downarrow}$ with the advantage that no compensating electric field is needed. Unlike in the $q$-filter, however, the beam structure is now strongly affected by the magnetic field. A TEM$_{00}$ Gaussian beam after passing through the nonuniform magnetic field of the device splits out into \textit{multi Gaussian-like} beams, each beam oscillating along the vector-potential minima. In particular, the incident Gaussian electron beam splits into two and three semi-Gaussian beams in quadrupole and hexapole magnetic fields, respectively. It is worth noting, however, that the multi Gaussian-like beam does recover its original Gaussian shape at certain free-space propagation distance, provided its phase distribution does not acquire sudden changes in the transverse plane. In this work, we introduce and numerically simulate two realistic configurations of spin-filter for electron beams, based on the new proposed device. Our numerical simulations confirms that a portion of electrons, usually small, remains polarized after passing through the device and can be easily separated form the rest of the beam by suitable apertures.
\section{Propagation of electron beams in an orthogonal uniform magnetic field}\label{sec:theory}
Let us assume that the electron beam moves along the $z$-direction perpendicular to a uniform magnetic field $\bm B=B_{0}(\cos{\theta},\sin{\theta},0)$, which lies in the $(x,y)$ transverse plane at angle $\theta$ with respect to the $x$-axis. As associated vector potential we may take is $\bm A=B_{0}\left(0,0,y\cos{\theta}-x\sin{\theta}\right)$. We assume a non relativistic electron beam, so that we can use Pauli's equation
\begin{equation}\label{eq:pauli}
   i\hbar\,\partial_t\tilde\psi = \left\{\frac{1}{2m}(-i\hbar\nabla-e \bm A)^2- \bm B \cdot \hat{\bm\mu}\right\}\tilde\psi,
\end{equation}
where $\tilde\psi$ is a two-component spinor and $\hat{\bm\mu}=\frac{1}{2}g\mu_B\hat{\bm\sigma}$ is the electron magnetic moment -- $\mu_B=\hbar e/2m$ is Bohr's magneton, $g$ is the electron $g$-factor, and $\hat{\bm\sigma}=(\hat\sigma_x,\hat\sigma_y,\hat\sigma_z)$ is Pauli's vector, respectively. We assume a paraxial beam with average linear momentum $p_c$ and average energy $E_c=p_c^2/2m$, so that $\tilde\psi(x,y,z,t)=\exp[i\hbar^{-1}(p_c z-E_c t)]\tilde u(x,y,z)$, with $\tilde u(x,y,z)$ slow-envelope spinor field~\cite{bliokh07}. Inserting this \textit{ansatz} into Eq.~(\ref{eq:pauli}) and neglecting the second derivatives of $\tilde u$ with respect to $z$, we obtain the paraxial Pauli equation
\begin{widetext}
\begin{equation}\label{eq:utilde}
   \left\{2ik_c\partial_z + \nabla^2_\perp +2k_c\,\frac{e}{\hbar}\, A - \frac{e^2}{\hbar^2}A^2+ \frac{2m}{\hbar^2}
   \bm B\cdot\hat{\bm\mu}\right\}\tilde u(x,y,z) = 0,
\end{equation}
\end{widetext}
where $\perp$ stands for the transverse coordinate, and $k_c=p_c/\hbar$ is a central de Broglie wave-vector.

Equation (\ref{eq:utilde}) is solved with initial Cauchy data at $z=0$, $\tilde u(r,\phi,0)=\tilde a\exp(-r^2/w_0^2)$ corresponding to a Gaussian beam of width $w_0$ in the cylindrical coordinates of $(r,\phi,z)$. The constant spinor $\tilde a=(a_1,a_2)$ describes the polarization state $\ket{\psi}=a_1\ket{\!\uparrow}+a_2\ket{\!\downarrow}$ of the input beam in the $\ket{\!\uparrow},\ket{\!\downarrow}$ basis where the spin is aligned parallel or antiparallel to the beam propagation direction, respectively. We assume the normalization $|a_1|^2+|a_2|^2=1$. A straightforward calculation shows that the required solution of the paraxial Pauli equation is given by
\begin{equation}\label{eq:utildesol}
   \tilde u(r,\phi,z)=G(r,\phi,z)\hat M(z) \tilde a,
\end{equation}
where $\hat M(z)$ is a matrix given by
\begin{equation}\label{eq:M}
  \hat M(z) = \begin{pmatrix}
             \cos\frac{2\pi z}{\Lambda_1} & i e^{-i\theta}\sin\frac{2\pi z}{\Lambda_1}\\
             i e^{i\theta}\sin\frac{2\pi z}{\Lambda_1} & \cos\frac{2\pi z}{\Lambda_1}
           \end{pmatrix},
\end{equation}
with $\Lambda_1=\frac{4\pi \hbar^2 k_c }{m g \mu_B B_0}$. The matrix $\hat M(z)$ accounts for the action of the magnetic field on the particle spin. The action on the electron motion is described in Eq.~(\ref{eq:utildesol}) by the Gaussian-Coherent factor $G(r,\phi,z)$ given by
\begin{equation}\label{eq:G}
   G(r,\phi,z) = \sqrt{\frac{-k_c\, z_R}{\pi\, q_\|(z)q_\bot(z)}}\,e^{i k_c\left(\frac{f_g(r,\phi)}{2q_\|(z)}+\frac{f_c(r,\phi,z)}{2q_\bot(z)}+\frac{z}{2}\right)}
\end{equation}
with
\begin{widetext}
\begin{eqnarray}\label{eq:f}
f_g(r,\phi)&=&r^2\cos^2(\theta-\phi),\\ \nonumber
f_c(r,\phi,z)&=&i\left(\frac{\pi}{\Lambda}\right)\left(\frac{\Lambda}{\pi}+r \sin{(\theta-\phi)}\right)\left(2i\left(\frac{\Lambda}{\pi}\right)^2+\left(\frac{\pi}{\Lambda}\right)z_R\,\left(q_\bot(z)+i z_R\cos{\left(\frac{\pi z}{\Lambda}\right)}\right)\left(\frac{\Lambda}{\pi}+r\sin{\left(\theta-\phi\right)}\right)\right) \cr
&+&\cos{\left(\frac{\pi z}{\Lambda}\right)}\left(2\left(\frac{\Lambda}{\pi}\right)^2+r \sin{\left(\theta-\phi\right)}\left(2\left(\frac{\Lambda}{\pi}\right)+r\sin{\left(\theta-\phi\right)}\right)\right).
\end{eqnarray}
\end{widetext}
The complex curvature radii of the Gaussian-Coherent factor $G(r,\phi,z)$ are given by

\begin{eqnarray}
   q_\|(z) &=& z - i z_R \label{eq:q1}\\ \label{eq:eq1}
   q_\bot(z) &=& \left(\frac{\Lambda}{\pi}\right) \sin{\left(\frac{\pi z}{\Lambda}\right)}-i z_R\cos{\left(\frac{\pi z}{\Lambda}\right)},  \label{eq:q2}
\end{eqnarray}
where $z_R=\frac{1}{2}k_c w_0^2$, $\Lambda=\frac{\pi\hbar k_c}{e B_0}$. From Eq.~(\ref{eq:M}) we see that, the beam spin state oscillates during propagation with spatial period $\Lambda_1$.
\begin{figure}[t]
   \centering
   \includegraphics[width=8.5cm]{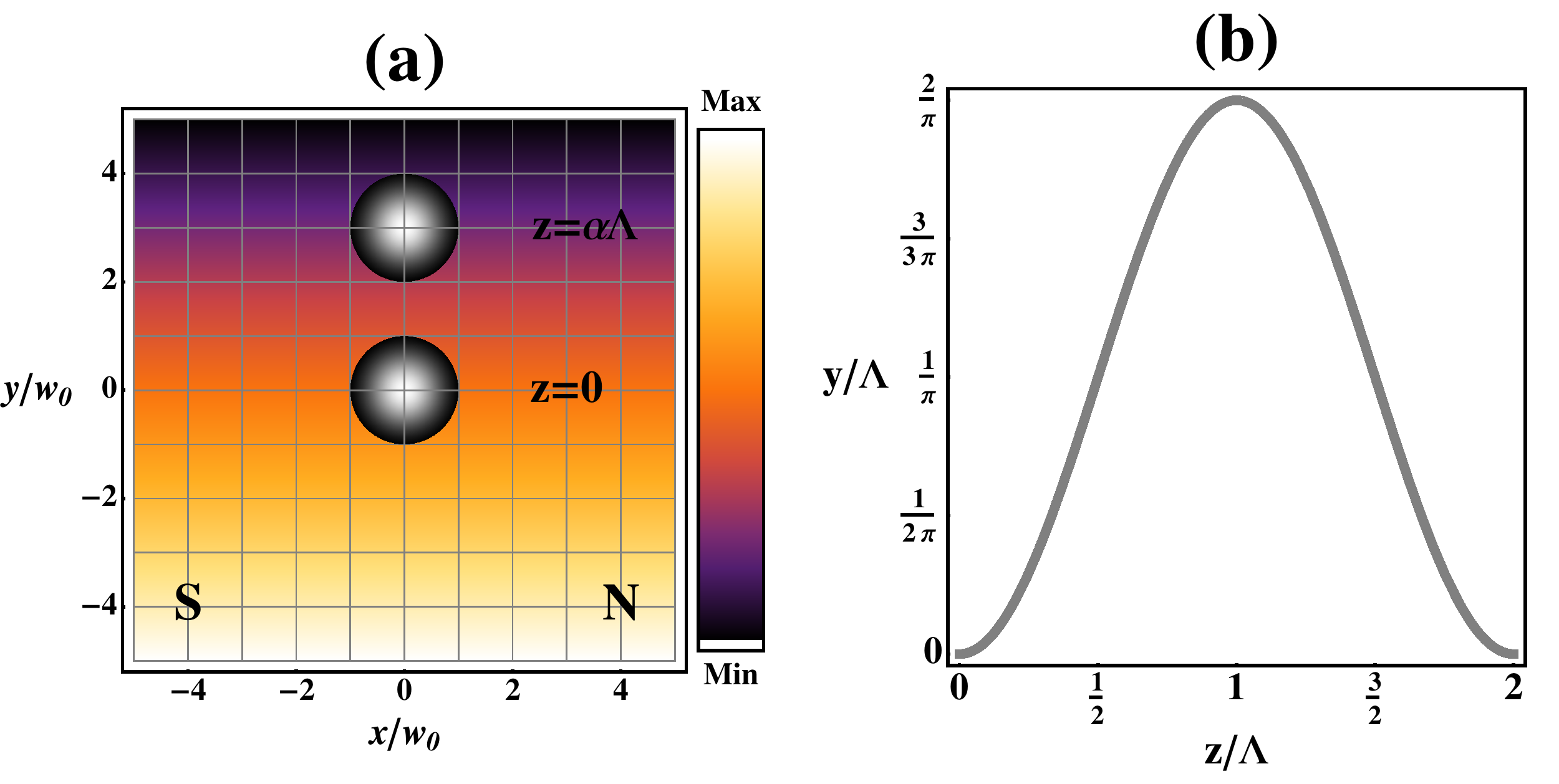}
   \caption{(a) Cross section of an electron gaussian beam propagating in a uniform magnetic field along the $x$-axis. During propagation the beam starts to move up in the positive $y$-direction, orthogonal to the magnetic field. The figure shows the beam position at $z=\alpha\Lambda$ with $\alpha=4.4\times10^{-3}$. The sidebar shows the strength of vector potential in false-color. (b) $y$-displacement of the beam center as function of the $z$ coordinate. The oscillation is sinusoidal and recovers its transverse position and shape at planes $z=n\Lambda$ ($n$ integer). The simulation was performed for an electron beam having energy $E_c=100$~KeV, waist $w_0=10\mu$m, in a magnetic field of strength $B_0=3.5$~mT. }
   \label{fig:uniform-magnetic}
\end{figure}
Moreover, unlike in the Wien-filter, which does not affect the beam mean direction, now the beam oscillates perpendicularly to the magnetic field $\bm B$ towards the minima of the vector potential with spatial period $2\Lambda=(g/4)\Lambda_1\simeq\Lambda_1/2\propto B_0^{-1}$. Figure \ref{fig:uniform-magnetic}-(a) shows the beam intensity profiles at two different $z$-planes; at the entrance plane $z=0$ (central spot) and at $z=\alpha\Lambda$ with $\alpha=0.44\%$ (upper spot). The electron trajectory in $yz$-plane, orthogonal to magnetic field, is shown in Figure \ref{fig:uniform-magnetic}-(b): the electron beam follows a sinusoidal oscillation with spatial period $2\Lambda$ and amplitude $2\Lambda/\pi$.
\section{Propagation of electron beams in an orthogonal nonuniform magnetic field possessing a specific topological charge}
\begin{figure}[t]
   \centering
   \includegraphics[width=8.5cm]{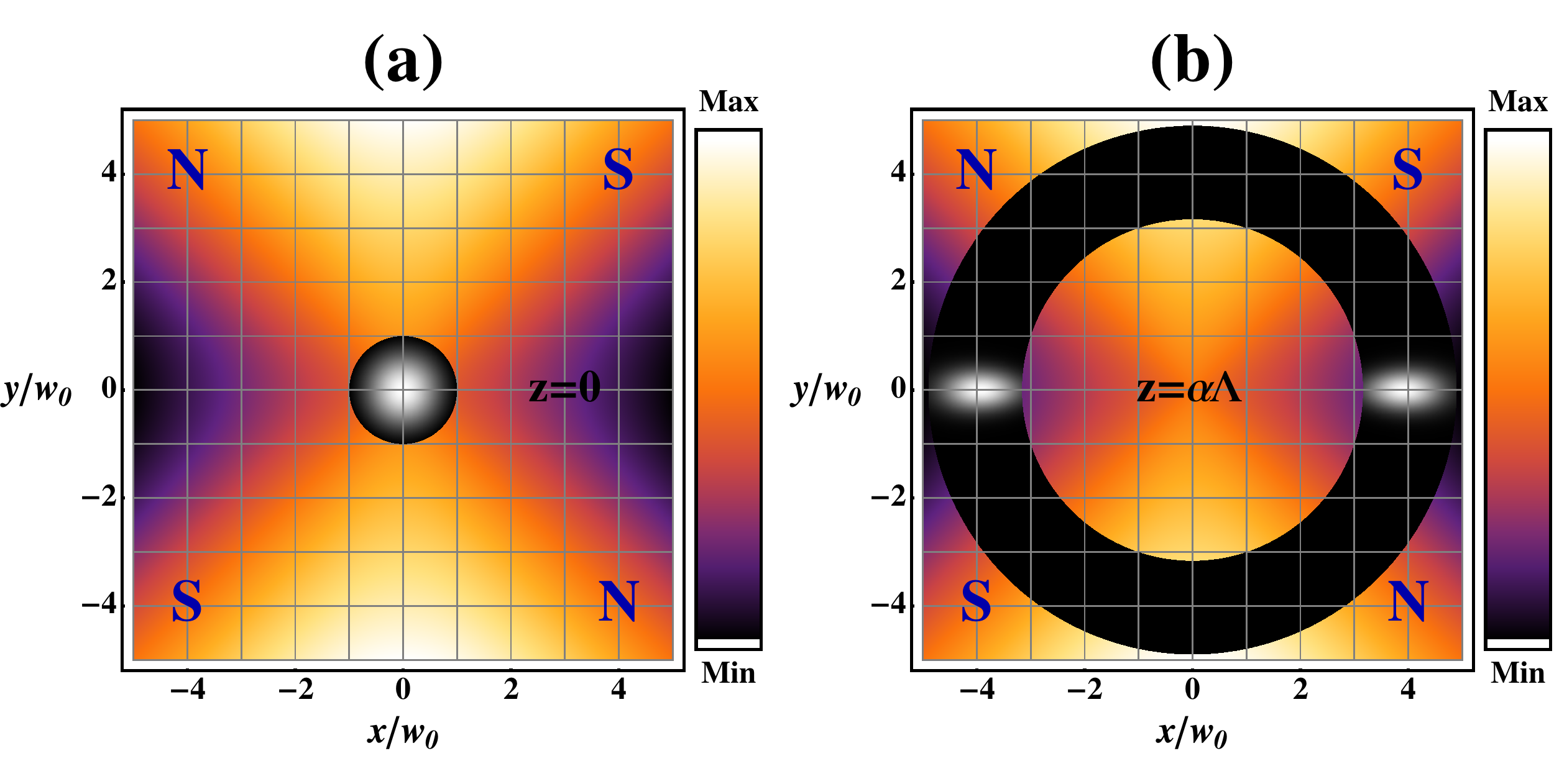}
   \caption{Propagation of an electron gaussian beam through a quadrupole ($q=-1$ and $\beta=-\pi/2$) magnetic field; (a) at the quadrupole pupil (b) after a propagation distance where is given by $z=\alpha \Lambda$. As shown in (b), the beam splits out into two different astigmatic quasi-Gaussian beams along vector-potential minima inside the quadrupole, along $x$-axis. The bar side shows the strength of vector potential in a false-color, and simulation was performed for $\alpha=4.4\times10^{-3}$.}
   \label{fig:quadrupoles}
\end{figure}

\begin{figure}[h]
   \centering
   \includegraphics[width=8.5cm]{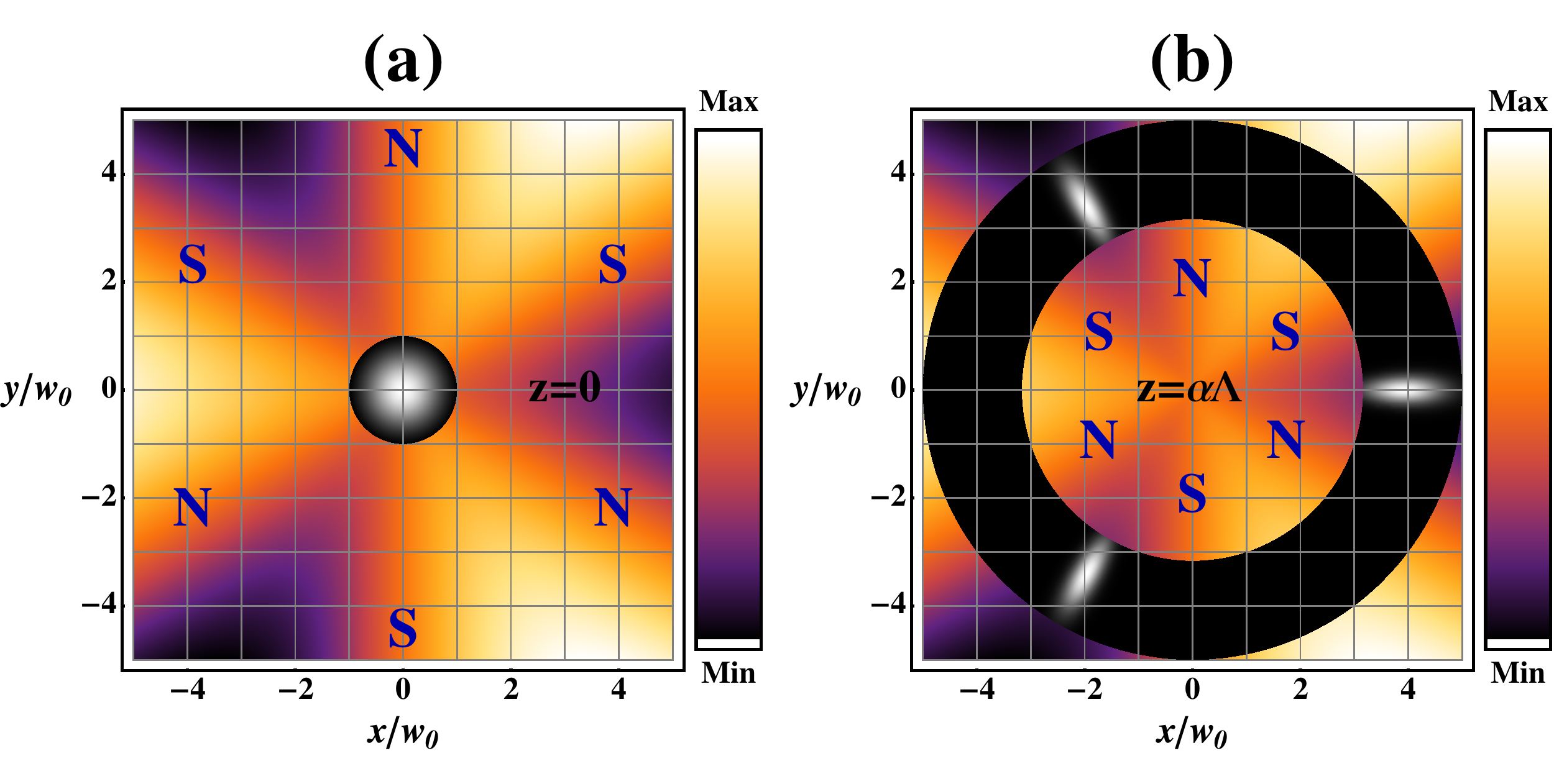}
   \caption{Propagation of an electron gaussian beam through a hexapoles ($q=-2$ and $\beta=-\pi/2$) magnetic field; (a) at the pupil of hexapoles (b) after a propagation distance where is given by $z=\alpha \Lambda$. As shown, during propagation the beam splits out into three different astigmatic quasi-Gaussian beams along vector-potential minima. The bar side shows the strength of vector potential in a false-color, and simulation is been performed for $\alpha=4.4\times10^{-3}$.}
   \label{fig:hexapole}
\end{figure}

Equation (\ref{eq:utildesol}) represents an exact solution of the beam paraxial equation, with an explicit boundary condition, for a uniform constant magnetic field at angle $\theta$ with respect to the $x$-axis. If the angle $\theta=\theta(x,y)$ changes slowly in the transverse plane, we may assume that the solution (\ref{eq:utildesol}) is still approximately valid. This Geometric Optics Approximation (GOA) is quite accurate in the present case, since the electron beam wavelength in a typical Transmission Electron Microscope (TEM) is in the range of tens of picometers, while $\theta$ changes over length of several microns. Within this slowly varying approximation, the effect of a nonuniform magnetic field is obtained simply by replacing $\theta$ with $\theta(x,y)$ in Eqs.~(\ref{eq:M}), (\ref{eq:G}), and (\ref{eq:f}). We assume singular space distribution of the magnetic field where $\theta(x,y)=\theta(r,\phi)$ is given by
\begin{equation}\label{eq:theta}
  \theta(\phi) = q\phi +\beta,
\end{equation}
where $\phi=\arctan{\left(y/x\right)}$ is the azimuthal angle in the beam transverse plane and $\beta$ is a constant angle, which defines the inclination on the $x$-axis. Finally $q$ is an integer which fixes the topological charge of the singular magnetic field distribution. Such magnetic structures can be generated in practice by multipolar lenses (for negative charges $q$) or by a set of appropriate longitudinal currents at origin (for positive charge $q$).  Inserting Eq.~(\ref{eq:theta}) into Eqs.~(\ref{eq:M}), (\ref{eq:f}), yields
\begin{equation}\label{eq:M1}
\hat M(z) = \begin{pmatrix}
             \cos\frac{2\pi z}{\Lambda_1} & i e^{-iq\phi}e^{-i\beta}\sin\frac{2\pi z}{\Lambda_1}\\
             i e^{iq\phi}e^{i\beta}\sin\frac{2\pi z}{\Lambda_1} & \cos\frac{2\pi z}{\Lambda_1}
           \end{pmatrix}
\end{equation}
and
\begin{widetext}
\begin{eqnarray}\label{eq:f1}
f_g(r,\phi)&=&r^2\cos^2((q-1)\phi+\beta),\\ \nonumber
f_c(r,\phi,z)&=&i\left(\frac{\pi}{\Lambda}\right)\left(\frac{\Lambda}{\pi}+r \sin{\left((q-1)\phi+\beta\right)}\right)\left(2i\left(\frac{\Lambda}{\pi}\right)^2+\left(\frac{\pi}{\Lambda}\right)z_R\,\left(q_\bot(z)+i z_R\cos{\left(\frac{\pi z}{\Lambda}\right)}\right)\left(\frac{\Lambda}{\pi}+r\sin{\left((q-1)\phi+\beta\right)}\right)\right) \cr
&+&\cos{\left(\frac{\pi z}{\Lambda}\right)}\left(2\left(\frac{\Lambda}{\pi}\right)^2+r \sin{\left((q-1)\phi+\beta\right)}\left(2\left(\frac{\Lambda}{\pi}\right)+r\sin{\left((q-1)\phi+\beta\right)}\right)\right).
\end{eqnarray}
\end{widetext}
Equation (\ref{eq:M1}) shows that in passing through a multipoles magnetic field of length $L$, a fraction $|\eta|^2=\left|\sin\frac{2\pi L}{\Lambda_1}\right|^2$ of the electrons in the beam flip their spin and acquire a phase factor $\exp(\pm i q\phi)$ accordingly if the initial spin were up $\ket{\!\uparrow}$ or down $\ket{\!\downarrow}$, respectively. The rest of the electrons, i.e., $1-|\eta|^2$, pass through without changing their initial spin state. When $L=\Lambda_1/2$ (mod $2\pi$), all electrons of the beam emerge with their spin reversed and acquire the above mentioned phase factor, which means that an amount of $\pm\hbar q$ is added to their initial OAM value. In this case the Spin-To-OAM Conversion (STOC) process is complete and we say that the device is \qo{\textit{tuned}}~\cite{karimi12}. Tuning can be made by acting on the strength of the magnetic field or changing the device length. The STOC process is governed by the $\exp(\pm iq\phi)$ factors in Eq.~(\ref{eq:M1}), which is of geometrical origin~\cite{marrucci11}. As a consequence, the STOC process occurs even if the field amplitude $B_0$ is (slowly) dependent on the radial coordinate $r$ in the beam transverse plane. The capability of changing the electron OAM and of creating a correlation between OAM and spin are the main features of the non-uniform magnetic multipoles. Equation (\ref{eq:f1}) shows that the device alters deeply the transverse profile of the beam, which acquires a multi quasi-Gaussian shapes as shown in Figs.~\ref{fig:quadrupoles}--(b) and \ref{fig:hexapole}--(b) for quadrupole and hexapole magnetic field, respectively.
The Gaussian beam at the entrance after propagation through the magnetic field breaks up respectively into two and three astigmatic quasi-Gaussian shape beams for quadrupole and hexapole. The cleft beams number depends on the number of vector potential minima. Both the STOC and non-STOC part of the beam possess the same intensity profile; but the STOC part only acquires a helical phase structure according to the magnetic field topological charge. For a small beam distortion, it can be shown that in the far-field the multi gaussian-like beam with the helical structure assumes a doughnut shape, while the vortex free non-STOC beam assumes a Gaussian shape. \newline
\section{Fringe fields and its effect on the spin-filtering}
In practice, it is impossible to generate a completely transverse magnetic field. A further non-transverse magnetic fields known as the \textit{fringe fields} cannot be avoided. In this section, we examine the effect of the fringe fields on the spatial pattern distribution of electron beam when our proposed device is not tuned. A non-tuned device, based on its length and magnetic field strength, converts only a portion of the incoming beam flipping the electron longitudinal--spin state and gaining OAM -  the remaining part of the beam left unchanged. When the non--tuned device is applied to an unpolarized electron beam, the beam exits in a mixture state of spin \textit{up} and \textit{down} with opposite OAM values, given by
\begin{eqnarray}\label{eq:mixture}
	\ket{\psi}_{\mbox{STOC}}=\eta
	\left\{ \begin{array}{ll}
         \ket{\uparrow,-\ell} & \mbox{for spin $\ket{\downarrow}$ input}\\
        \ket{\downarrow,+\ell} & \mbox{for spin $\ket{\uparrow}$ input} 
        \end{array} \right.,
\end{eqnarray}
\begin{figure}[b]
   \centering
   \includegraphics[width=8.5cm]{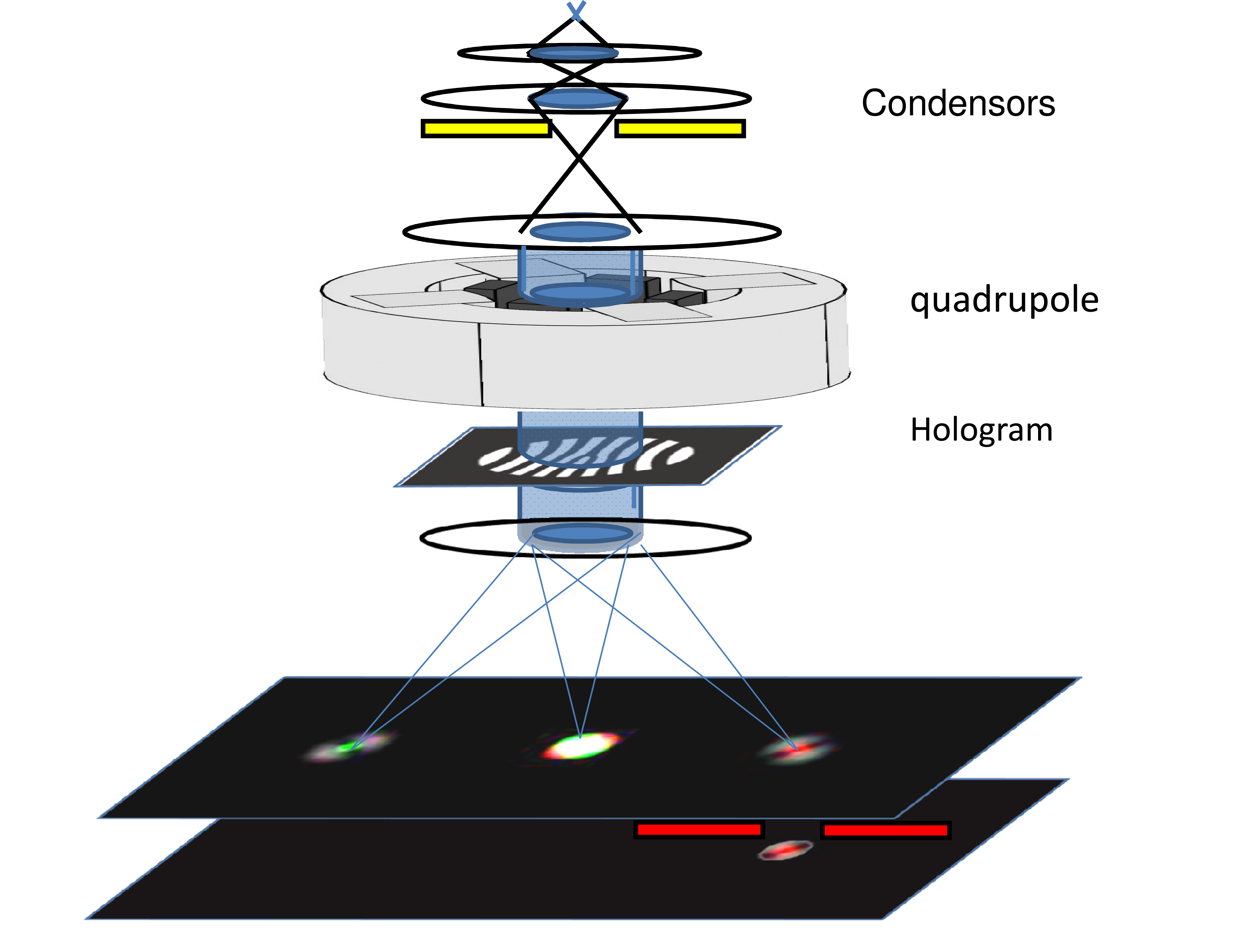}
   \caption{First proposed scheme to generate a spin-polarized electron beam based on space-variant magnetic fields.}
   \label{fig:setup1}
\end{figure}
where $\ket{\ell}$ stands for the OAM state given by topological charge of device, i.e., $\ell=q$, and $|\eta|^2=|\sin\frac{2\pi L}{\Lambda_1}|^2$ is the device STOC's efficiency. The main fringe fields appear at the entrance and the exit face of the device as nonuniform longitudinal magnetic fields. The interaction of longitudinal magnetic field with both spin and orbital angular momentum of electron beams has been recently theoretically investigated~\cite{greenshields12}. The longitudinal magnetic field introduces a phase rotation $\Phi(\ell,s)\propto B_z(\ell+g\,s)z$ on the beam, where $s=\pm1$ is the spin eigenvalues in unite of $\hbar/2$ corresponding to spin \textit{up} and \textit{down}, respectively, and $B_z$ is in general a function of transverse radial coordinate $r$.~\cite{greenshields12}. This phase rotation comes out from the Zeeman interaction. The longitudinal fringe field introduces a different phase change in each term of Eq.~(\ref{eq:mixture}) which is therefore changed into
\begin{eqnarray}\label{eq:mixt_fring}
	\ket{\psi}_{\mbox{STOC}}=\eta
	\left\{ \begin{array}{ll}
         e^{i\Phi(-\ell,1)}\ket{\uparrow,-\ell} & \mbox{for spin $\ket{\downarrow}$ input}\\
       	e^{i\Phi(\ell,-1)}\ket{\downarrow,+\ell} & \mbox{for spin $\ket{\uparrow}$ input}.
        \end{array} \right.\qquad
\end{eqnarray}
Since the fringe fields are nonuniform, in general, the phases in Eq.~(\ref{eq:mixt_fring}) are coordinate dependent and produce dual converging and diverging astigmatic effects. However, even in the presence of fringe fields, the two spin states of the emerging beam are still labeled by the two values $\pm\ell$ of OAM so that an OAM sorter can be used to separate the electrons according to their spin value. In the next section, we calculated numerically the efficiency of this way to obtain polarized electron beam by using a pitch-fork hologram with topological charge $\ell$ as OAM sorter. The final electrons state after an OAM sorter, then, is
\begin{eqnarray}\label{eq:mixt_fring1}
	\ket{\psi}_{\mbox{final}}=\eta
	\left\{ \begin{array}{ll}
         e^{i\Phi(-\ell,1)}\ket{\uparrow,0} & \mbox{for spin $\ket{\downarrow}$ input}\\
       	e^{i\Phi(\ell,-1)}\ket{\downarrow,+2\ell} & \mbox{for spin $\ket{\uparrow}$ down}.
        \end{array} \right.\qquad
\end{eqnarray}
The first term tends to recover the Gaussian shape, while the second term will have a doughnut shape in the far-field. Both of the non-STOC terms own OAM=$\ell$ since the hologram, i.e., OAM sorter, is spin independent. A spatial selector can be use, e.g., a pinhole, to select the central part, which has a uniform coherent spin \textit{up} state. The efficiency and purity of the spin filter depends on the the pinhole radius, which for an optical field has been discussed in~\cite{karimi09}. So, combination of this device with an OAM sorter yields an electron spin-filter that can be realistic since the multipolar magnetic magnets are available commercially as an aberration corrector for TEM as well as OAM sorters.
\section{Numerical simulation and technical discussions}
The GOA analytical solution presented in Section~\ref{sec:theory} might be used to obtain preliminary information on the STOC efficiency and beam intensity profile in the device. However, more details on the interaction of a space-variant magnetic field with an electron beam can be examined by implementing both a ray tracing technique and the spin--orbit interaction simultaneously. In order to overcome this issue, we developed a research--didactic software based on the \textit{multi-slice method} used in the electron microscopy, where Eq. (\ref{eq:utilde}) has been considered as a free particle motion under the force of a \qo{\textit{local potential}} of $A$, $A^2$ and Pauli terms. Therefore, one can evolve both temporal and $z$-dependence of the wave-function based on a well-know \textit{Dyson}--like decomposition, since both the operators of the scalar wave equation and local potential do not commute~\cite{cowley57,pozzi89}. The potential was divided and projected into multi-slices where the beam wave-function was constructed by a free-space propagation between each slices. After each iteration the wave-function turns into
\begin{figure}[t]
   \centering
   \includegraphics[width=8cm]{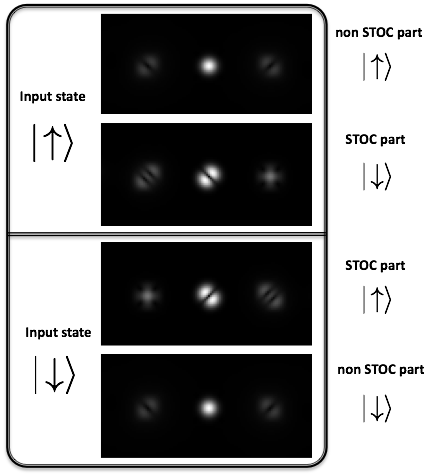}
   \caption{Simulated diffraction pattern of the beam generated by a quadrupole after passing through a pitch-fork hologram (right spot assumed to be the first order of diffraction). In a fraction of the beam the nonuniform magnetic field of the quadruple couples electron spin to OAM. Up and down subfigures inside each row show the non--converted and converted parts of an input 50-50 mixture of spin \textit{up} and \textit{down}, respectively. The electrons spin state in the first order of diffraction has been indicated at the right side of images.}
   \label{fig:simulations1}
\end{figure}
\begin{equation}\label{eq:multislice}
u(\mathbf{r}_\bot,z_{j+1})={\mathcal{K}}\otimes\,\left(e^{\left(\frac{i}{\hbar v} \int_{z_j}^{z_{j+1}}V(\mathbf{r}_\bot,\zeta)\, d\zeta\right)}\cdot u({\mathbf{r}_\bot,z_j})\right),
\end{equation}
where $v$ is the electrons velocity, $\otimes$ is the convolution with the wave-function inside parenthesis, $z_j$ and $V(\mathbf{r}_\bot,\zeta)$ stand for position of $j^{th}$ slice and the local potential (third, fourth and fifth term of Eq. (\ref{eq:utilde})), respectively (see Ref.~\cite{grillo:13} for more details). $\mathcal{K}$ is the Fresnel propagator between each adjoined two slices spaced by $\Delta z$, which is given by
\begin{equation}\label{eq:fresnel}
\mathcal{K}=\frac{-i k_c}{2\pi\Delta z}\,e^{\frac{i k_c}{2\Delta z}\left(x^2+y^2\right)}.
\end{equation}
\begin{figure}[t]
   \centering
   \includegraphics[width=8.5cm]{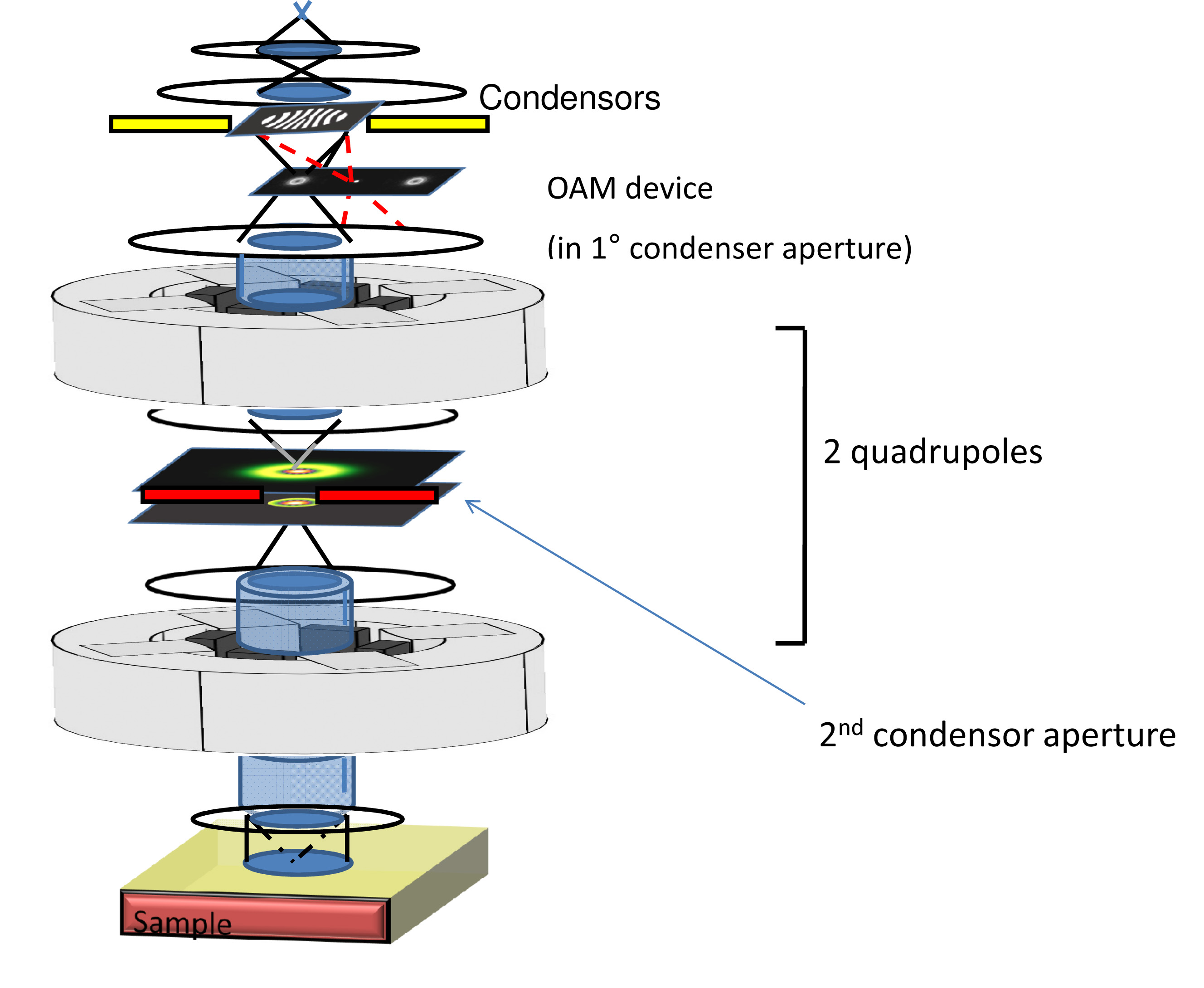}
   \caption{Second proposed scheme to generate a spin-polarized electron beam based on coupling with two magnetic quadruples (spherical aberration corrector). The second quadruple corrects the aberration induced by the first one. Nevertheless, the spin-to-orbit coupling efficiency after the second quadruple can be completely neglected.}
   \label{fig:setup2}
\end{figure}
Apart the simple concept, one may extend such a powerful algorithm to the relativistic case as well~\cite{rother08}. However, since electron microscopes work at mid-range energy, our simulation was carried out in the non-relativistic regime. We considered two possible configurations to generate a spin-polarized electron beam in an electron microscope.\newline\newline
\begin{figure}[t]
   \centering
   \includegraphics[width=8.5cm]{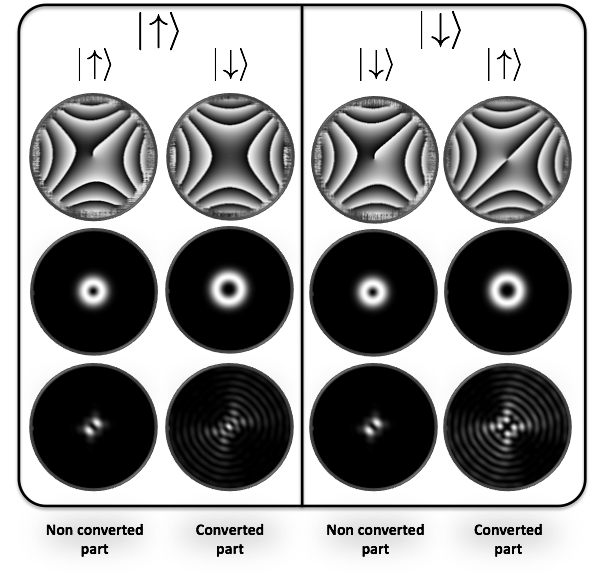}
   \caption{As the beam with OAM=$+1$ propagates inside the aberration corrector system (quadrupole - condenser - aperture and a rotated quadrupole), it sees an astigmatism effect introduced by the quadrupoles and, based on the time interaction, a portion of electron suffers spin-to-orbit conversion, second and fourth column. The first two rows show phase and intensity distributions of electron beam after the first quadrupole for a mixture of 50-50 spin \textit{up} and \textit{down}, respectively.  The last row shows the output intensity distribution for both converted and non-converted part of spin up and down after interacting with whole system. The simulation was made for an input electron beam size of 1$\mu$m, a quadrupole magnetic field of $0.1$mT at the beam waist radius and a device thickness of 10$\mu$m. This relatively small thickness was chosen to simplify and speed up the numerical calculation and to avoid too large phase deformations.}
   \label{fig:simulation2} 
\end{figure}
\noindent \textbf{(i)} A first scheme can be in principle adopted in many microscopes using a condenser stigmator. In this first scheme, a Gaussian beam is directed to a quadrupolar magnetic field, and then to a pitch-fork hologram as shown in Fig. (\ref{fig:setup1}). A lens condenser can be used to form the far-field image of the hologram. However, the intensity required for the magnetic field might make the operation difficult. It also turns out difficult to polarize electrons before the specimen, except in microscopes with two or even three condensers aperture planes.\newline
The simulated electron beam shapes after passing through the quadrupole and the pitch-fork hologram for spin \textit{up} and \textit{down} input are shown in Fig (\ref{fig:simulations1}). \newline
The first and second row in Fig.~(\ref{fig:simulations1}) show the far-field intensity pattern for input electrons with spin \textit{up} and \textit{down} state, respectively. The subrows are correspond to expected spin state for the first order of diffraction (left spots). The non-STOC part of both spin \textit{up} and \textit{down}, i.e., first and last rows of Fig.~(\ref{fig:simulations1}), forms a \textit{Hermite-Gaussian} shape of the first order, which is affected by the quadrupole's astigmatism and splits into two parts. Conversely the STOC parts are shaped differently depending on the initial spin state; one as the \textit{Hermite-Gaussian} of second order and the other one forms Gaussian beam bearing some astigmatic distortion. It can be seen that an aperture with appropriate size can be used to select the central spot only, which possesses opposite polarization. 
\newline
\noindent \textbf{(ii)} In the second case, a simplified scheme of a spherical aberration corrector has been considered, see Fig (\ref{fig:setup2}), where two quadrupoles with opposite polarization were coupled through two cylindrical lenses (transport lenses) having magnetic field along the propagation direction. Differently from the real device, we will assume that a limiting aperture can be added inside the corrector in correspondence of the focal plane of the transport lenses. Figure (\ref{fig:simulation2}) shows the simulated evolution of the wave-function in the system for both spin \textit{up} and \textit{down}. At the exit face of the first quadrupole the two polarizations are indistinguishable, but evolve to a different intensity distribution in the focal plane where an aperture selects the central part of the beam containing mainly the $\ket{\!\uparrow,0}$ electron state. The second quadruple, indeed, implies aberration correction into the selected $\ket{\!\uparrow,0}$ beam. It is worth noticing that the spin-to-orbit coupling, due to strength of magnetic field, after the second quadruple can be neglected.
\section{Conclusions}
We presented two possible practical devices to generate a spin polarized electron beam via a spin-to-orbit conversion in the presence of a nonuniform transverse magnetic field. The heart of the devices is a multipolar magnet generating a singular transverse magnetic structure with negative integer topological charge. The device action on a pure spin-polarized electron beam is spin-to-orbit conversion and the beam gains a nonuniform phase structure, defined by the magnetic field topological charge, where the sign is given by the input electron spin value. When the device is combined with an OAM sorter, spiral phase plate, hologram, or even longitudinal magnetic field, it can be used as a spin-filter to polarize the electron beam. However, because of the strong astigmatism, the efficiency of the devices proposed here are lower than the ones discussed in Ref~\cite{karimi12}.
\section{Acknowledgement}
V. G. would like to thank Prof. M. Heider and Dr. H. Mueller for useful discussion. E. K. and R. W. B. acknowledge the support of the Canada Excellence Research Chairs (CERC) program. E. S. acknowledges the financial support of the Future and Emerging Technologies (FET) programme within the Seventh Framework Programme for Research of the European Commission, under FET-Open grant number 255914-PHORBITECH.

\end{document}